\documentclass[preprint,prd,aps,showpacs,showkeys,nofootinbib]{revtex4-1}
\usepackage{amssymb}
\usepackage{amsmath}
\usepackage{graphicx}
\usepackage{subfigure}
\usepackage{float}
\usepackage{mathrsfs}

\usepackage{amsmath}
\usepackage{epstopdf}
\usepackage{dcolumn}
\usepackage{bm}
\usepackage{subfig}

\newcommand{\be}{\begin{eqnarray}}
\newcommand{\ee}{\end{eqnarray}}
\newcommand{\PP}{{\cal P}_{\cal R}}

\UseRawInputEncoding
\begin{document}


\title{NANOGrav Signal from double-inflection-point inflation and dark matter}

\author{Tie-Jun Gao}
\email{tjgao@xidian.edu.cn}
\affiliation{School of Physics and Optoelectronic Engineering, Xidian University, Xi'an 710071, China}

\begin{abstract}
The NANOGrav collaboration has published a suspected stochastic GW background signal in its analysis of 12.5 years PTA data, so in this work, we investigate the possibility to explain the signal by the inflationary models with double-inflection-point. We calculate the energy spectrum of GWs induced by scalar perturbations, and show that the curve lies in the $2\sigma$ region of the NANOGrav constraints. In addition, we analyze the reheating process and dark matter production by assuming that the inflaton is coupled with the SM Higgs boson and singlet fermionic dark matter field. We discuss the radiative stability of the inflationary potential under one-loop corrections, calculate the reheating temperature, the dark matter production, and constraints the coupling parameters using the bounds of BBN, Lyman-$\alpha$, etc.

\end{abstract}

\maketitle

\section{Introduction \label{sec:intr}}
Recently, the North American Nanohertz Observatory for Gravitational Waves (NANOGrav) has published its 12.5 years observation data of pulsar timing array(PTA), where strong evidence of a stochastic process, which can be explained by the stochastic gravitational waves (GWs) with a power-law spectrum $\Omega_{GW}\propto f^{5-\gamma}$ at a reference frequency of $f_{yr}\simeq3.1\times10^{-8}$Hz, with the exponent $5-\gamma\in (-1.5, 0.5)$ at $1\sigma$ confidence level\cite{Arzoumanian:2020vkk,Vaskonen:2020lbd,Kohri:2020qqd,DeLuca:2020agl,Kawasaki:2021ycf,Vagnozzi:2020gtf,Domenech:2020ers}.

It has been pointed out in several literatures that if the power spectrum of scalar perturbations has a large peak at low scales, then when the perturbations corresponding to the peak renters the horizon during the radiation-dominated period, it will induce GWs, which is sizable to be detectable by experiments in near future\cite{ref8,ref9,ref10,ref11,ref12,ref13,ref14,ref15,ref16,ref171,ref172,ref173,ref174,ref175,Domenech:2021ztg}.
Such enhancement of the power spectrum can be achieved in the ultra-slow-roll phase near the inflection point in some inflationary models~\cite{ref18,ref19,ref20,ref21,ref22,Gao:2019kto}, and similar models have been discussed in many literatures ~\cite{ref23,ref231,ref24},or in the framework of string theory~\cite{ref27,ref28,ref29,ref30} etc. However, In the previous models, the potential contains single inflection point, and the inflation will last about more than 30 e-folding numbers before the inflection point. Thus near the inflection point, the peak of the power spectrum will induced GWs around millihertz, which couldn't explain the NANOGrav result around nanohertz. So in this paper, motivated from the framework of effective field theory, we consider a polynomial potential model with double-inflection-point. In such a model, the inflection point at CMB scales can make the predictions consistent with the  2018 data\cite{planck} and last about 20 e-folding numbers, thus when the inflaton meets the second inflection point, it will induce GWs with the peak around nanohertz, which can explain the  NANOGrav signal.

After inflation ends, the inflaton will oscillate around the minimum of the potential and decay into relativistic particles, which will reheating the universe. So in this paper, we assume that the inflaton can decay into the standard model(SM) Higgs boson or decay into singlet fermions beyond SM, which can be a component of dark matter(DM). In order to ensure that the effect of the added coupling terms do not affect the inflationary dynamics at the CMB scale,and do not affect the GW energy spectrum, we discuss the radiative corrections at one-loop order using the Coleman-Weinberg (CW) formalism~\cite{CW,Weinberg:1973am}. We also calculate the reheating temperature and constraint the model parameters using BBN\cite{BBN1,BBN2,BBN3,BBN4,BBN5}, Lyman-$\alpha$~\cite{lya1,lya2}, etc.  We also analysis the dark matter production,  and show that the main way to produce dark matter is the direct decay of inflatons.

The paper is organized as follows.
In the next section, we setup the inflationary model with double-inflection-point using a scalar potential with polynomial form. In Sec.3, we analyze the  inflation dynamics of the model and calculate the power spectrum numerically. In Sec.4, we calculate the energy spectrum of the induced GWs and compared the results with NANOGrav and some planned experiments. In Sec.4, we assume that after inflation, the inflaton will decay into SM Higgs or singlet fermionic  field and analyze the reheating temperature. We also calculate the effect of one-loop corrections of the coupling terms in Sec.5. In Sec. 6, we discuss the dark matter production and calculate the relic density.
The last section is devoted to summary.


\section{The double-inflection-point model}
 In this section, we consider a scalar potential with polynomial form, which can generate an inflationary model with double-inflection-point. Such a polynomial can be derived from the effective field theory with a cutoff scale $\Lambda$ \cite{Bhaumik:2019tvl,Enqvist:2003gh,Burgess:2009ea,Marchesano:2014mla,ref37,ref38}
\begin{equation}
V_{\mathrm{eff}}(\phi)=\sum_{n=0} \frac{b_{n}}{n !}\left(\frac{\phi}{\Lambda}\right)^{n}.
\end{equation}
To obtain a inflationary potential containing double-inflection-point, we truncate the above polynomial to the sixth order, and redefine the parameters as
\begin{eqnarray}
V(\phi)=V_{0}\left[c_{2}\left(\frac{\phi}{\Lambda}\right)^{2}+c_{3}\left(\frac{\phi}{\Lambda}\right)^{3}+c_{4}\left(\frac{\phi}{\Lambda}\right)^{4}+c_{5}\left(\frac{\phi}{\Lambda}\right)^{5}+\left(\frac{\phi}{\Lambda}\right)^{6}\right],
\end{eqnarray}
where the overall factor $V_0$ can be constrained
 by the amplitude of scalar perturbations $A_s$, and we have omit the constant and first-order terms of the polynomial
so that the potential and it's first-order derivative vanish at the origin. By tuning the four dimensionless parameters $c_{2-5}$ one can obtain a potential with two inflection points.

For the purpose of discussion, we assume that the two inflection points are located at $\phi_i (i=1,2)$, where the first and second derivatives of $V$ vanish,i.e. $V'(\phi_i)=0$ and  $V''(\phi_i)=0$. This condition yields the following relationship between $c_{2-5}$ and $\phi_i$
\begin{eqnarray}
c_2=\frac{3}{\Lambda^4} \phi_1^2 \phi_2^2,c_3=  \frac{-4}{\Lambda^3} \left(\phi_1^2 \phi_2+\phi_1 \phi_2^2\right),c_4=\frac{3}{2\Lambda^2} \left(\phi_1^2+4 \phi_1 \phi_2+\phi_2^2\right),c_5= \frac{-12}{5\Lambda}  (\phi_1+\phi_2).
\end{eqnarray}
However, in order to obtain a reasonable model, the inflection points of the potential are not strict, so we introduce two additional parameters $\alpha_i$ to represent the small deviation. Then the scalar potential can be written in the following form
\begin{eqnarray}
V(\phi)=&&\frac{V_{0}}{\Lambda^6}\Big(3 \phi_1^2 \phi_2^2\phi^{2}-4 \left(\phi_1^2 \phi_2+\phi_1 \phi_2^2\right)(1+\alpha_1)\phi^{3}\nonumber\\
&&+\frac{3}{2} \left(\phi_1^2+4 \phi_1 \phi_2+\phi_2^2\right)\phi^{4}-\frac{12}{5}  (\phi_1+\phi_2)(1+\alpha_2)\phi^{5}+\phi^{6}\Big).
\end{eqnarray}
For some parameter spaces, the model is both consistent with the CMB observational constraints and interprets the NANOGrav signal. For instance, we choose
\begin{eqnarray}
&&V_0=2.157\times10^{-13}M_p^4,\ \ \  \Lambda=M_p,\nonumber\\
&&\phi_1=0.9095875M_p,\ \ \ \ \ \ \ \alpha_1= 8.5654\times10^{-5},\nonumber\\
&& \phi_2=2.10081M_p, \ \ \ \ \ \ \ \ \ \ \alpha_2= -4.0172\times10^{-5},
\end{eqnarray}
which corresponds to the parameters $c_{2-5}$ are
\begin{eqnarray}
c_2=10.9543,~ c_3= -23.0119,~ c_4= 19.3264,~ c_5= -7.22466.
\end{eqnarray}
In Fig. 1 we draw the corresponding potential, which contains double-inflection-point.
\begin{figure}\small

  \centering
   \includegraphics[width=4in]{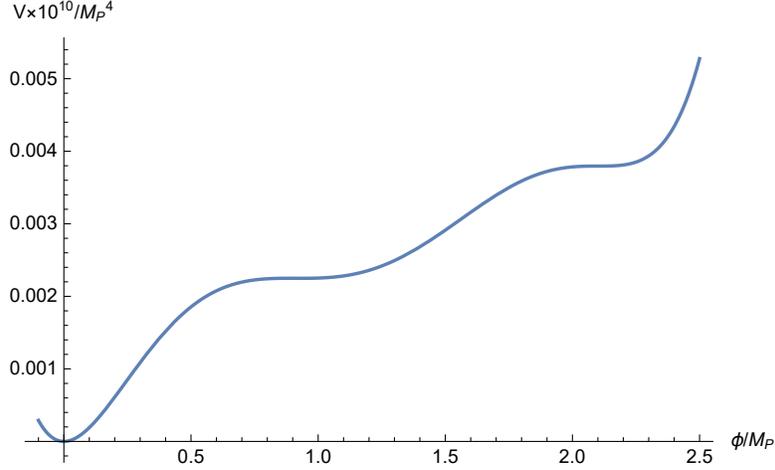}
     \caption{The scalar potential with parameter set (5).}
    \label{fig1}
\end{figure}
As we will see below, the inflation starts near the first inflection point at high scale and generates a nearly scale-invariant power spectrum, which is in good agreement with the CMB observations, and when the inflaton rolls near the second inflection point at low scale, it will go through an ultra-slow-roll phase, which will last about 35 e-folding numbers and resulting in a large peak in the power spectrum,  which will induces gravitational waves consistent with the NANOGrav data.

\section{Inflation dynamics}
In this section, we will discuss the dynamics of inflation. Since there is an ultra-slow-roll stage in the process of inflation, so in order to calculate the power spectrum more accurately, we use the Hubble slow roll parameters defined below~\cite{ref40,ref41,ref42,ref43,ref44,ref45,ref46}
\begin{eqnarray}
&&\epsilon_H=-\frac{\dot{H}}{H^2},\nonumber\\
&&\eta_H=-\frac{\ddot{H}}{2H\dot{H}}=\epsilon_H-\frac{1}{2}\frac{d\ln\epsilon_H}{dN_e},
\end{eqnarray}
with dots represent derivatives with respect to the cosmic time, and $N_e$ is the $e$-folding numbers. The corresponding curves of $\epsilon_H$ and $\eta_H$ with respect to $N_e$ are shown in Fig.2,  and the evolution of Hubble parameter with respect to $N_e$ are shown in Fig.3.
\begin{figure}\small

  \centering
   \includegraphics[width=4in]{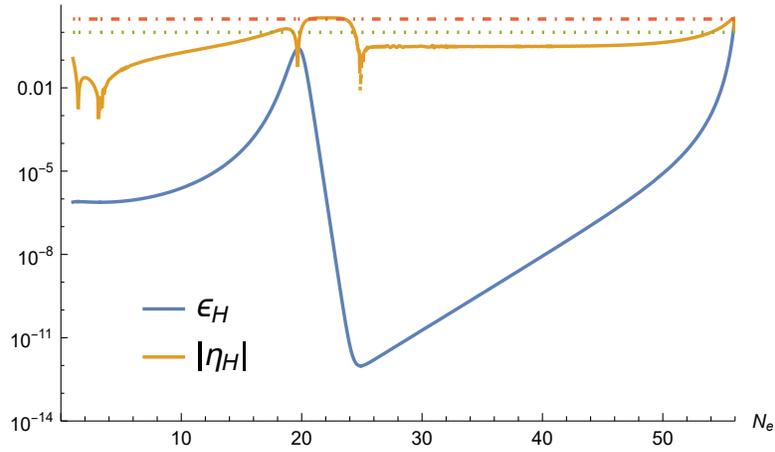}
     \caption{The Hubble slow-roll parameters $\epsilon_H$ and $\eta_H$ as functions of the e-folding number $N_e$. The green and red dashed lines represent 1 and 3, respectively.}
    \label{fig1}
\end{figure}
\begin{figure}\small

  \centering
   \includegraphics[width=4in]{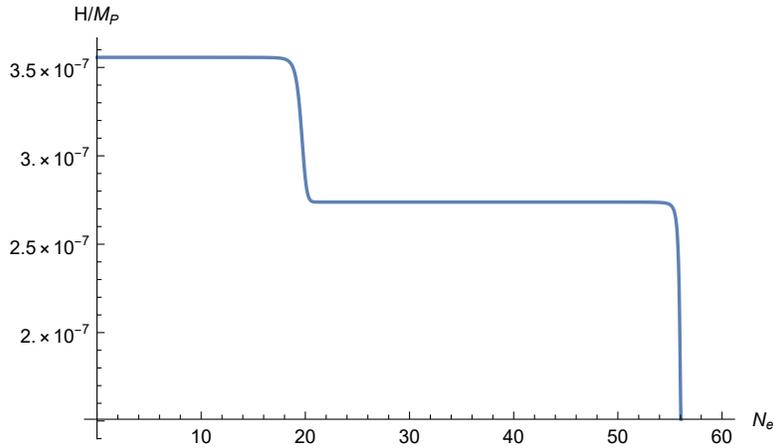}
     \caption{The Hubble parameters as functions of the e-folding number $N_e$}
    \label{fig1}
\end{figure}

As we can see in the figure, near the second inflection point, the slow-roll parameter $|\eta_H|>3$, which implies that inflation undergoes an ultra-slow-roll process. Correspondingly, the curve of the slow-roll parameter $\epsilon_H$ appears a deep valley lasting about 30 e-folding numbers, which will make the power spectrum appear a large peak. In order to calculate the power spectrum at the ultra-slow-roll process, the slow-roll approximation is no longer applicable, so the Mukhanov-Sasaki(MS) equation of mode function $u_k$, must be solved strictly~\cite{ref21}
\begin{eqnarray}
&&\frac{d^2u_k}{d\tau^2}+\Big(k^2-\frac{1}{z}\frac{d^2z}{d\tau^2}\Big)u_k=0,
\end{eqnarray}
with $z\equiv\frac{a}{\mathcal{H}}\frac{d\phi}{d\tau}$, and $\tau$ denotes the conformal time.
And the initial condition is taken to be the Bunch-Davies type\cite{Bunch:1978yq}
\begin{eqnarray}
&&u_k\rightarrow \frac{e^{-ik\tau}}{\sqrt{2k}},\;\;  \text{as}\;\; \frac{k}{aH}\rightarrow\infty.
\end{eqnarray}
Then the power spectrum are calculated by
\begin{eqnarray}
&&\PP=\frac{k^3}{2\pi^2}\Big|\frac{u_k}{z}\Big|^2_{k\ll aH}.
\end{eqnarray}
We show the numerical result(blue line) and the approximate results(orange line) of the scalar power spectrum in Fig.4. And the constraints to the primordial power spectrum from $\mu$-distortion of CMB are also show there.
\begin{figure}\small

  \centering
   \includegraphics[width=4in]{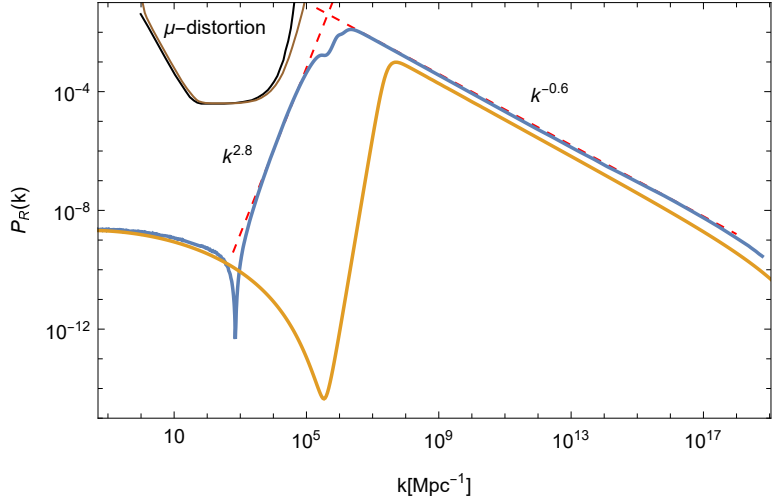}
     \caption{The power spectrum of scalar perturbations with the parameter set(5). And the black and brown lines show the  upper bound from $\mu$-distortion for a delta function power spectrum and for the steepest growth $k^4$ power spectrum, respectively\cite{Byrnes:2018txb}.}
    \label{fig1}
\end{figure}
We can see that the ultra-slow-roll behavior near the inflection point leads to a large peak in the power spectrum, and the peak value is about seven orders of magnitude higher than the power spectrum of the CMB scale. We will see in the next section that the perturbations corresponding to the peak will induce experimentally detected GWs after re-entering the horizon, and can explain the signal of NANOGrav.

In order to verify that the model agrees with the constraints of Planck experiment on CMB scale, we estimate the corresponding scalar spectral index and the tensor-to-scalar ratio, which can be expressed using $\epsilon_H$ and $\eta_H$ at the leading order as
\begin{eqnarray}
&&n_s=1-4\epsilon_H+2\eta_H,\nonumber\\
&&r=16\epsilon_H.
\end{eqnarray}
The numerical results are $n_s=0.9672$, $r=1.22\times10^{-5}$, and the amplitude of the primordial curvature perturbations $A_s$ and the e-folding numbers during inflation $N_e$ are  $\ln(10^{10}A_s)=3.0444$, $N_e=56.6$. The results are all consistent with the observation constraints from Planck 2018, which are $n_s=0.9649\pm0.0042$, $r<0.064$ and  $\ln(10^{10}A_s)=3.044\pm0.014$\cite{planck}.

\section{Induced Gravitational waves\label{sec:gw}}
In the following, we will numerical calculate the second order GWs induced by scalar perturbations using the power spectrum obtained in the province subsection.
When the scalar perturbation re-enters the horizon, it will induce second-order GWs\cite{ref31,ref32,ref33,ref34,ref35,ref36}, and the corresponding GW energy spectrum can be expressed by the tensor power spectrum as
\begin{equation}
\label{eq:OmegaGW}
\Omega_{\mathrm{GW}}(\tau, k) = \frac{1}{24}\left(\frac{k}{\mathcal{H}}\right)^{2} \overline{{\mathcal P}_{h}(\tau, k)}\,,
\end{equation}
where the overline denotes the oscillation average among several wavelengths.
Using the Green's function method and considering that $\mathcal{H}=1/\tau$ in the radiation dominant period, the above energy spectrum can be calculated by the scalar power spectrum as following~\cite{ref13}
\be
&& \Omega_{\mathrm{GW}}(\tau,k)
 = \dfrac{1}{12}\int_{0}^{\infty} dv \int_{|1-v|}^{1+v} du \left(\frac{4 v^{2}-\left(1+v^{2}-u^{2}\right)^{2}}{4 u v}\right)^{2} \PP (k u) \PP (k v) \nonumber \\
&& \quad \left(\frac{3}{4 u^{3} v^{3} }\right)^{2}\left(u^{2}+v^{2}-3\right)^{2} \nonumber \\
&& \quad \Big\{\Big[-4 u v+\left(u^{2}+v^{2}-3\right) \ln \left|\frac{3-(u+v)^{2}}{3-(u-v)^{2}}\right|\Big]^{2}+\left[\pi\left(u^{2}+v^{2}-3\right) \Theta(u+v-\sqrt{3})\right]^{2} \Big\},
\label{eq:final}
\ee
where $u $ and $v $  are two dimensionless variables.
Finally, the energy density spectrum of GWs today $\Omega_{\mathrm{GW}, 0}$ is calculated by\cite{ref33}
\begin{equation}
\label{eq:Omegaappr}
\Omega_{\mathrm{GW}, 0}=0.83 \left(\frac{g_{*}}{g_{*,p}}\right)^{-1 / 3} \Omega_{r, 0} \Omega_{\mathrm{GW}},
\end{equation}
with $\Omega_{r, 0}\simeq9.1\times10^{-5}$ is the energy density fraction of radiation at present,
$g_{*}$ and $g_{*,p}$ denote the effective number of degrees of freedom for energy density today and at the horizon crossing, respectively.

Combine the numerical result of scalar power spectrum $\PP$ obtained in the previous subsection, we numerically calculate the energy spectrum of induced GWs and show it in Fig.5, with the horizontal axis is the frequency at present
\begin{equation}
\label{eq:Omegaappr}
f \approx 0.03 \mathrm{Hz} \frac{k}{2\times 10^7 \mathrm{pc}^{-1}}.
\end{equation}
And the upper curves are the sensitivity curves of some planned
GW detectors~\cite{ref52,ref53,ref54,ref55,ref56,ref24}.
\begin{figure}\small

  \centering
   \includegraphics[width=4in]{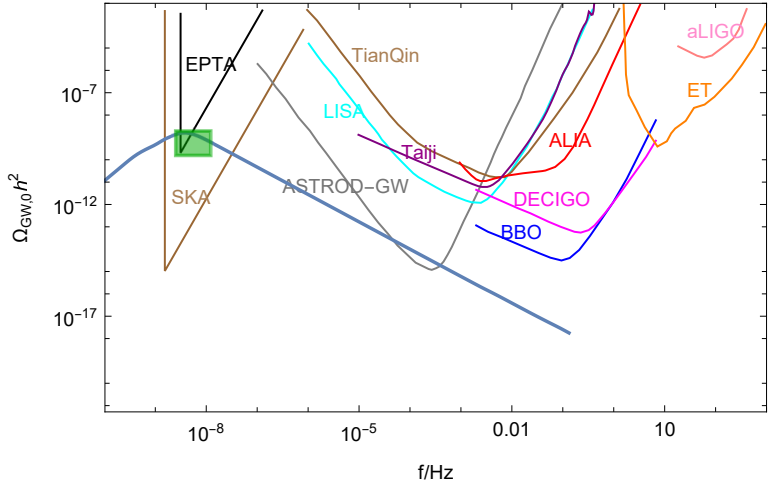}
     \caption{Energy spectrum of the induced GWs at the present time
predicted by the polynomial model for parameter set (5).
The curves in the upper part are the expected sensitivity curves of the European Pulsar Timing Array (EPTA), Square Kilometer Array (SKA),Laser Interferometer Space Antenna (LISA),Taiji, TianQin,Astrodynamical Space
Test of Relativity using Optical-GW detector (ASTROD-GW), Advanced Laser Interferometer Antenna (ALIA), Big Bang Observer (BBO), Deci-hertz Interferometer GW Observatory (DECIGO),Einstein Telescope (ET),  Advanced LIGO (aLIGO), respectively. These sensitivity curves are taken from Ref.~\cite{ref52,ref53,ref54,ref55,ref56,ref24} The green region  show the  $2 \sigma$ confidence level of the NANOGrav results with the tilt of $5-\gamma=0$\cite{Arzoumanian:2020vkk}.}
    \label{fig1}
\end{figure}

Recently, the NANOGrav collaboration has published its 12.5-year data of PTA, which indicate a signal which can be explained by the stochastic GWs with a power-law spectrum around $f_{yr}\simeq3.1\times10^{-8}$Hz,
\begin{equation}
\Omega_{GW}(f)=\frac{2\pi^2f_{yr}^2}{3H_0^2}A^2_{GWB}\Big(\frac{f}{f_{yr}}\Big)^{5-\gamma},
\end{equation}
where $H_0\equiv100 h$ km/s/Mpc, and $5-\gamma\in(-1.5,0.5)$ at $1\sigma$ confidence level\cite{Arzoumanian:2020vkk,Vaskonen:2020lbd,Kohri:2020qqd,DeLuca:2020agl,Kawasaki:2021ycf,Vagnozzi:2020gtf,Domenech:2020ers}. The observed GWs for $5-\gamma=0$ with $2 \sigma$ uncertainty on $A_{GWB}$ are also show in Fig.5.

We can see that the frequencies of the spectrum of GWs cover from nanohertz to millihertz, and the maximum is at the frequency $f=3.97\times10^{-9}Hz$, which is within the frequency range of SKA and EPTA. The  the spectrum of induced GWs with frequencies around nanohertz  lies in the $2\sigma$ region of the NANOGrav constraints, so it can explain the NANOGrav signals. And around millihertz, the energy spectrum curves lies above the expected sensitivity curves of ASTROD-GW, so it can be tested by the observation in near future.

\section{Reheating \label{sec:infl}}
After inflation ends, the inflaton rolls down the potential and then oscillates around the minimum, and the energy of inflaton will transfers to other degree of freedoms and raises the temperature of the universe. This is known as the reheating period.
Following Reference~\cite{Drees:2021wgd,Bernal:2021qrl,Ghoshal:2022jeo}, we consider the inflaton decays into SM Higgs boson or decays into a Dirac fermion $\chi$ through trilinear coupling, which can be a candidate of DM. The additional terms in the Lagrangian density has the following form
\begin{equation}
\mathcal{L}=i \bar{\chi} \gamma^\mu \partial_\mu \chi-m_\chi \bar{\chi} \chi-y_\chi \phi \bar{\chi} \chi-\lambda_{12} \phi H^{\dagger}H-\frac{1}{2}\lambda_{22} \phi^2 H^{\dagger}H,
\end{equation}
where $m_\chi$ is the mass of the DM, the coupling coefficient $y_\chi$, $\lambda_{22}$ are dimensionless and $\lambda_{12}$ has a dimension of mass.
The associated decay widths are
\begin{equation}
\begin{aligned}
\Gamma_{\phi \rightarrow H^{\dagger} H} & \simeq \frac{\lambda_{12}^2}{8 \pi m_\phi}, \\
\Gamma_{\phi \rightarrow \bar{\chi} \chi} & \simeq \frac{y_\chi^2 m_\phi}{8 \pi},
\end{aligned}
\end{equation}
with $m_\phi^2=\frac{\partial^2V}{\partial\phi^2}|_{\phi=0}$ is the inflaton mass during reheating, and we have assumed that the mass of $\chi$ and $H$ are much smaller then $m_\phi$.
In order to fit the relic density of photon numbers and hadronic numbers today, the decay width to the SM Higgs boson should be much greater than the decay width to the fermionic DM, that is the total decay width of the inflaton be approximated as $\Gamma\equiv\Gamma_{\phi\rightarrow H^{\dag}H}+\Gamma_{\phi\rightarrow \bar{\chi}\chi}\simeq\Gamma_{\phi\rightarrow H^{\dag}H}$.
Therefore branching ratio of DM production can be calculated as follows
\begin{equation}
\textrm{Br}\equiv\frac{\Gamma_{\phi\rightarrow \bar{\chi}\chi}}{\Gamma_{\phi\rightarrow H^{\dag}H}+\Gamma_{\phi\rightarrow \bar{\chi}\chi}}\simeq\frac{\Gamma_{\phi\rightarrow \bar{\chi}\chi}}{\Gamma_{\phi\rightarrow H^{\dag}H}}\simeq m^2_\phi(\frac{y_\chi^2}{\lambda_{12}^2}).
\end{equation}

During reheating, when the Hubble parameter becomes small enough, the energy loss due to the decay of the inflaton is greater than the energy loss due to the expansion of the universe, the corresponding temperature when $H=\frac{2}{3}\Gamma$ is defined as the reheating temperature $T_{rh}$. In the instantaneous decay approximation the reheating temperature can be calculated as~\cite{Bernal:2021qrl}
\begin{equation}
T_{rh}=\sqrt{\frac{2}{\pi}}\left(\frac{10}{g_{*}}\right)^{1 / 4} \sqrt{M_P} \sqrt{\Gamma},
\end{equation}
where $\Gamma$ is the total decay width of the inflaton, and $g_{*}=106.75$.
And the maximum temperature during reheating can be calculated by~\cite{Tmax1,Tmax2,Tmax3}
\begin{equation}
T_{\max }=\Gamma^{1 / 4}\left(\frac{60}{g_{*} \pi^2}\right)^{1 / 4}\left(\frac{3}{8}\right)^{2 / 5} H_I^{1 / 4} M_P^{1 / 2},
\end{equation}
where $H_I$ is the Hubble parameter at the beginning of reheating. In our model, according to Fig.3, we use the value around the second inflection point.

According to the restriction of BBN, the reheating temperature $T_{rh}$ should be greater than $4$MeV\cite{BBN1,BBN2,BBN3,BBN4,BBN5}. Moreover, the Planck 2018 give an upper limit on the inflation scale $H_I\leq2.5\times10^{-5}M_p$~\cite{planck}, which would allows an upper limit on the reheating temperature $T_{rh}\leq 7\times10^{15}$GeV. So we can give the limit of the coupling parameter $\lambda_{12}$ according to equation (20)
\begin{equation}
2.7451\times10^{-23}\leq\frac{\lambda_{12}}{M_p}\leq4.8039\times10^{-5}.
\end{equation}

In addition, we estimate the value of $T_{\max }/T_{rh}$
\begin{equation}
\frac{T_{\max }}{T_{rh}}=\left(\frac{3}{8}\right)^{2/5}\left(\frac{3M_p}{\pi}\sqrt{\frac{10}{g_*}}\frac{\mathcal{H}_I}{T_{rh}^2}\right)^{1/4},
\end{equation}
and show the allowed ranges in Fig.6. Where the constraints of red dashed line is from the BBN $T_{rh}>4MeV$ and the stability in the next section, $T_{rh}<4.43\times10^{12}$GeV by the upper bound of $\lambda_{12}$ in Eq.(29).
\begin{figure}\small

  \centering
   \includegraphics[width=4in]{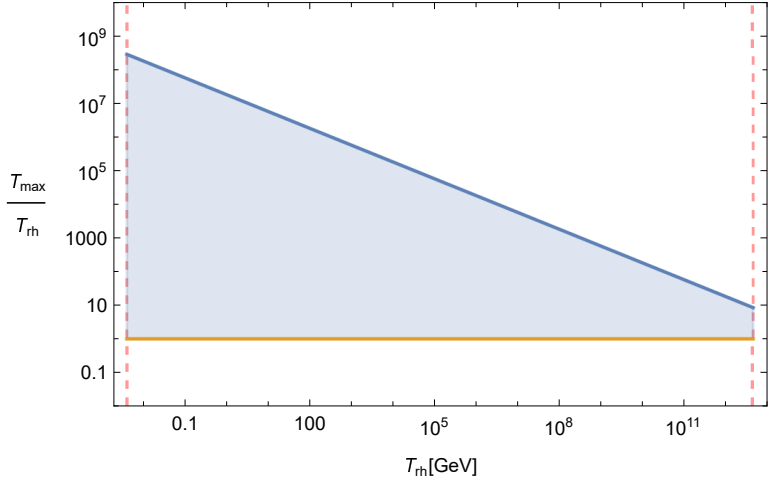}
     \caption{The allowed ranges of the ratio $T_{\max }/T_{rh}$. }
    \label{fig1}
\end{figure}

\section{Radiative Corrections and Stability \label{sec:infl}}
In order to ensure that the coupling terms added in the discussion of the reheating process in the previous section do not affect the inflationary dynamics at the CMB scale, and do not affect the generation of GWs at the second inflection point, thus do not affect the parameter space of the model, we analysis the stability of the inflation potential, calculate the one-loop CW correction of these coupling terms to the inflationary potential and restrict the coupling parameters.

The one-loop CW correction is~\cite{CW,Weinberg:1973am}
\begin{equation}
\Delta V=\sum_j \frac{g_j}{64 \pi^2}(-1)^{2 s_j} \tilde{m}_j^4\left[\ln \left(\frac{\widetilde{m}_j^2}{\mu^2}\right)-\frac{3}{2}\right],
\end{equation}
and the corresponding first and second derivative of $V_{CW}$ are
\begin{equation}
\begin{aligned}
&V_{\mathrm{CW}}^{\prime}=\sum_j \frac{g_j}{32 \pi^2}(-1)^{2 s_j} \tilde{m}_j^2\left(\widetilde{m}_j^2\right)^{\prime}\left[\ln \left(\frac{\widetilde{m}_j^2}{\mu^2}\right)-1\right], \\
&V_{\mathrm{CW}}^{\prime \prime}=\sum_j \frac{g_j}{32 \pi^2}(-1)^{2 s_j}\left\{\left[\left(\left(\widetilde{m}_j^2\right)^{\prime}\right)^2+\widetilde{m}_j^2\left(\widetilde{m}_j^2\right)^{\prime \prime}\right] \ln \left(\frac{\widetilde{m}_j^2}{\mu^2}\right)-\widetilde{m}_j^2\left(\widetilde{m}_j^2\right)^{\prime \prime}\right\},
\end{aligned}
\end{equation}
where the index $j$ is summing over three fields, the inflaton, the Higgs $H$ and the fermion $\chi$, and the spin $s_j$ are $s_H=0,s_{\chi}=1/2,s_{\phi}=0$,the number of degrees of freedom of the fields are $g_H=4,g_{\chi}=4,g_{\phi}=0$, respectively. The renormalization scale $\mu$ is taken as $\phi_0$.
$\tilde{m}_j$ are field-dependent masses, and in our model they are given by
\begin{equation}
\begin{aligned}
\widetilde{m}_\phi^2(\phi)&=\frac{6 V_0 }{\Lambda ^6}\Big(5 \phi ^4-8 (\alpha_2+1) \phi ^3 (\phi_1+\phi_2)+3 \phi ^2 \left(\phi_1^2+4 \phi_1 \phi_2+\phi_2^2\right)\\
&-4 (\alpha_1+1) \phi  \phi_1 \phi_2 (\phi_1+\phi_2)+\phi_1^2 \phi_2^2\Big), \\
\widetilde{m}_\chi^2(\phi) & =\left(m_\chi+y_{\chi} \phi\right)^2, \\
\widetilde{m}_{H}^2(\phi) & =m_{H}^2+\lambda_{12} \phi.
\end{aligned}
\end{equation}

At the high scale inflection point $\phi_2$, the first and second derivatives of inflation potential are
\begin{equation}
\begin{aligned}
 V'({\phi _2}) &=  - \frac{{12{V_0}}}{{{\Lambda ^6}}}{\phi _2}^3({\phi _1} + {\phi _2})({\alpha _1}{\phi _1} + {\alpha _2}{\phi _2}), \\
 V''({\phi _2}) &=  - \frac{{24{V_0}}}{{{\Lambda ^6}}}{\phi _2}^2({\phi _1} + {\phi _2})({\alpha _1}{\phi _1} + 2{\alpha _2}{\phi _2}), \\
\end{aligned}
\end{equation}
and at such point, the derivatives of the one-loop CW correction are
\begin{equation}
\begin{aligned}
{{V}_{CW}'}({\phi _2}) &  =\frac{{{\lambda_{12}^2}{\phi _2}(\ln \left( {\frac{\lambda_{12}}{{{\phi _2}}}} \right) - 1) - 2{\phi _2}^3y_\chi ^4(\ln \left( {y_\chi ^2} \right) - 1)}}{{8{\pi ^2}}} \\
   & + \frac{1}{{{\pi ^2}{\Lambda ^{12}}}}9{V_0}^2{\phi _2}^3\Big(\ln \left( { - \frac{{24{V_0}({\phi _1} + {\phi _2})({\alpha _1}{\phi _1} + 2{\alpha _2}{\phi _3})}}{{{\Lambda ^6}}}} \right) - 1\Big)({\phi _1} + {\phi _2})({\alpha _1}{\phi _1} + 2{\alpha _2}{\phi _2}) \\
   &\times((2{\alpha _1} - 1){\phi _1}^2 + 2(1 + {\alpha _1} + 6{\alpha _2}){\phi _1}{\phi _2} + (12{\alpha _2} - 1){\phi _2}^2),\\
   {{V}_{CW}''}({\phi _2})& = \frac{{{\lambda_{12}^2}\ln \left( {\frac{\lambda_{12}}{{{\phi _2}}}} \right) + 2{\phi _2}^2y_\chi ^4(1 - 3\ln \left( {y_\chi ^2} \right))}}{{8{\pi ^2}}}\\
  & + \frac{1}{{4{\pi ^2}{\Lambda ^{12}}}}9{V_0}^2{\phi _2}^2\Big(3{((1 - 2{\alpha _1}){\phi _1}^2 - 2(1 + {\alpha _1} + 6{\alpha _2}){\phi _1}{\phi _2} + (1 - 12{\alpha _2}){\phi _2}^2)^2} \\
 & - 6({\phi _1} + {\phi _2})({\alpha _1}{\phi _1} + 2{\alpha _2}{\phi _2})({\phi _1}^2 + (3 - 8{\alpha _2}){\phi _2}^2 - 4{\phi _1}({\phi _2} + 2{\alpha _2}{\phi _2})) \\
  &+ 2{((1 - 2{\alpha _1}){\phi _1}^2 - 2(1 + {\alpha _1} + 6{\alpha _2}){\phi _1}{\phi _2} + (1 - 12{\alpha _2}){\phi _2}^2)^2}\\
  &\times({\rm{ln}}\left( { - \frac{{24V_0({\phi _1} + {\phi _2})({\alpha _1}{\phi _1} + 2{\alpha _2}{\phi _2})}}{{{\Lambda ^6}}}} \right) - \frac{3}{2}) \\
 & - 6({\phi _1} + {\phi _2})({\alpha _1}{\phi _1} + 2{\alpha _2}{\phi _2})({\phi _1}^2 + (3 - 8{\alpha _2}){\phi _2}^2 - 4{\phi _1}({\phi _2} + 2{\alpha _2}{\phi _2}))\\
 &\times(2\ln \left( { - \frac{{24{V_0}({\phi _1} + {\phi _2})({\alpha _1}{\phi _1} + 2{\alpha _2}{\phi _2})}}{{{\Lambda ^6}}}} \right) - 3)\Big). \\
\end{aligned}
\end{equation}
In order to make sure the addition coupling terms do not effect the inflation dynamics, we need to make sure that the terms of $y_\chi^4$ and $\lambda_{12}^2$ are much smaller than the tree-level results(27), which will give the following restrictions on the parameters spaces
\begin{equation}
\begin{aligned}
y_\chi<1.00687\times10^{-4},\\
\lambda_{12}<3.03996\times10^{-8}M_p,\\
y_\chi<1.8046\times10^{-4},\\
\lambda_{12}<1.73968\times10^{-7}M_p.
\end{aligned}
\end{equation}
In addition, in our model we must also need to ensure that the addition coupled terms will not affect the generation of GWs, that is, the corresponding terms of $y_\chi^4$ and $\lambda_{12}^2$ in the one-loop CW corrections at the low scale inflection point
\begin{equation}
\begin{aligned}
{{V}_{CW}'}({\phi _2}) & = \frac{{{\lambda _{12}}^2{\phi _1}(\ln \left( {\frac{{{\lambda _{12}}}}{{{\phi _1}}}} \right) - 1) - 2{\phi _1}^3(\ln \left( {y_\chi ^2} \right) - 1)y_\chi ^4}}{{8{\pi ^2}}} \\
  &+ \frac{1}{{{\pi ^2}{\Lambda ^{12}}}}9{V_0}^2{\phi _1}^3(\ln \left( { - \frac{{24{V_0}({\phi _1} + {\phi _2})(2{\alpha _2}{\phi _1} + {\alpha _1}{\phi _2})}}{{{\Lambda ^6}}}} \right) - 1)({\phi _1} + {\phi _2})(2{\alpha _2}{\phi _1} + {\alpha _1}{\phi _2}) \\
  &\times ((12{\alpha _2} - 1){\phi _1}^2 + 2(1 + {\alpha _1} + 6{\alpha _2}){\phi _1}{\phi _2} + (2{\alpha _1} - 1){\phi _2}^2), \\
   {{V}_{CW}''}({\phi _2})& =\frac{{({\lambda _{12}}^2\ln \left( {\frac{{{\lambda _{12}}}}{{{\phi _1}}}} \right) - 2{\phi _1}^2(3\ln \left( {y_\chi ^2} \right) - 1)y_\chi ^4)}}{{8{\pi ^2}}}\\
   &+\frac{1}{{4{\pi ^2}{\Lambda ^{12}}}}9{V_0}^2{\phi _1}^2\Big(3{((1 - 12{\alpha _2}){\phi _1}^2 - 2(1 + {\alpha _1} + 6{\alpha _2}){\phi _1}{\phi _2} + (1 - 2{\alpha _1}){\phi _2}^2)^2} \\
 & + 6({\phi _1} + {\phi _2})(2{\alpha _2}{\phi _1} + {\alpha _1}{\phi _2})(( - 3 + 8{\alpha _2}){\phi _1}^2 - {\phi _2}^2 + 4{\phi _1}({\phi _2} + 2{\alpha _2}{\phi _2}))\\
  &+ 2{((1 - 12{\alpha _2}){\phi _1}^2 - 2(1 + {\alpha _1} + 6{\alpha _2}){\phi _1}{\phi _2} + (1 - 2{\alpha _1}){\phi _2}^2)^2} \\
  &\times (\ln \left( { - \frac{{24{V_0}({\phi _1} + {\phi _2})(2{\alpha _2}{\phi _1} + {\alpha _1}{\phi _2})}}{{{\Lambda ^6}}}} \right) - \frac{3}{2}) \\
  &+ 6({\phi _1} + {\phi _2})(2{\alpha _2}{\phi _1} + {\alpha _1}{\phi _2})(( - 3 + 8{\alpha _2}){\phi _1}^2 - {\phi _2}^2 + 4{\phi _1}({\phi _2} + 2{\alpha _2}{\phi _2})) \\
  &\times (2\ln \left( { - \frac{{24{V_0}({\phi _1} + {\phi _2})(2{\alpha _2}{\phi _1} + {\alpha _1}{\phi _2})}}{{{\Lambda ^6}}}} \right) - 3)\Big),
   \end{aligned}
\end{equation}
are much smaller than the terms of tree-level
\begin{equation}
\begin{aligned}
 V'({\phi _1}) &=  - \frac{{12{V_0}}}{{{\Lambda ^6}}}{\phi _1}^3({\phi _1} + {\phi _2})({\alpha _2}{\phi _1} + {\alpha _1}{\phi _2}), \\
 V''({\phi _1}) &=  - \frac{{24{V_0}}}{{{\Lambda ^6}}}{\phi _1}^2({\phi _1} + {\phi _2})(2{\alpha _2}{\phi _1} + {\alpha _1}{\phi _2}), \\
\end{aligned}
\end{equation}
which will give the following restrictions
\begin{equation}
\begin{aligned}
y_\chi<2.2307\times10^{-4},\\
\lambda_{12}<6.46594\times10^{-8}M_p,\\
y_\chi<1.88144\times10^{-4},\\
\lambda_{12}<8.18895\times10^{-8}M_p.
\end{aligned}
\end{equation}
Combine 29 and 32, we get the upper bounds are $y_\chi<1.00687\times10^{-4}$ and $\lambda_{12}<3.03996\times10^{-8}M_p$. Plugging the upper limit of $\lambda_{12}$ into (20), we can  give an upper limit on the reheating temperature $T_{rh}<4.43\times10^{12}$GeV.




\section{Dark Matter Production and Relic Density \label{sec:infl}}
In this section, we will study the dark matter production during reheating.
Combining the Boltzmann equation of DM number density $n_\chi$ and the Friedman equation, and considering that during $T_{rh}<T<T_{max}$ the energy density is dominated by the inflaton, then we can obtain the following relationship between the comoving number density $N=n_\chi a^3$ of DM and the reheating temperature $T_{rh}$ as\cite{Bernal:2021qrl}
\begin{equation}
\frac{d N}{d T}=-\frac{8}{\pi} \sqrt{\frac{10}{g_{*}}} \frac{M_p T_{\mathrm{rh}}^{10}}{T^{13}} a^3\left(T_{\mathrm{rh}}\right) \gamma,
\end{equation}
where $a$ is the scale factor and $\gamma$ is the density of DM production rate. And then the DM yield $Y\equiv n_\chi/s$ can be expressed as
\begin{equation}
\frac{d Y}{d T}=-\frac{135}{2 \pi^3 g_{*,s}} \sqrt{\frac{10}{g_{*}}} \frac{M_p}{T^6} \gamma,
\end{equation}
where $s\equiv\frac{2\pi^2}{45}g_{*,s}T^3$is the entropy density at temperature $T$, and $g_{*, s}$  is the
number of relativistic degrees of freedom contributing to the SM entropy~\cite{SM entropy}. In addition, it is  worth to note that in order to consistent with observations of DM energy density , the present day DM yield is fixed by\cite{Bernal:2021qrl}
\begin{equation}
m_\chi Y_0\simeq4.3\times10^{-10}\textrm{GeV}.
\end{equation}

After inflation, the DM can produced by the direct decay of inflatons, the 2-to-2 scattering of inflatons and the 2-to-2 scattering of SM particles.

The main way of dark matter production is the direct decay of inflatons, in this case, the DM production rate density is
\begin{equation}
\gamma=2 \textrm{Br} \Gamma \frac{\rho}{m}.
\end{equation}
Using Eq. (34), we can get that the corresponding DM yield in this case is
\begin{equation}
Y_0\simeq \frac{3}{\pi} \frac{g_{*}}{g_{*,s}} \sqrt{\frac{10}{g_{*}}} \frac{M_p \Gamma}{m_\phi T_{\mathrm{rh}}} \mathrm{Br} 
\\\simeq1.163\times10^{-2}M_p \frac{y_\chi^2}{T_{rh}},
\end{equation}
In the above equation we have assume that $g_{*,s}=g_*$. Combined with Eq.(35), we obtain the conditions if the inflatons decay constitutes the whole DM abundance, and show the  allowed range of the coupling coefficient $y_\chi$ in Fig. 7.
\begin{figure}\small

  \centering
   \includegraphics[width=4in]{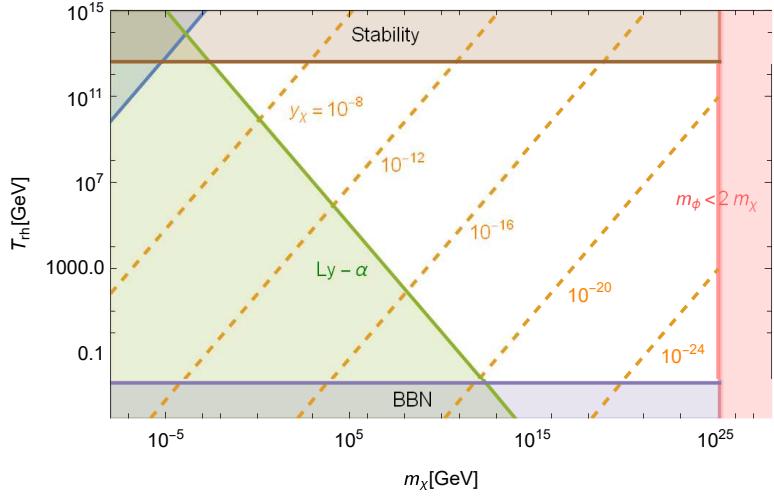}
     \caption{ The allowed range of coupling parameters $y_\chi$ when the direct decay of the inflaton produces the whole DM.}
    \label{fig1}
\end{figure}
Where the constraints of purple region is from the BBN, $T_{rh}>4MeV$, the brown and blue region are from the upper bound of $T_{rh}<4.43\times10^{12}$Gev by the discussion of stability in Section 6, the green region is from the $Lyman-\alpha$ bound $\frac{m_\chi}{keV}\geq\frac{2m_\phi}{T_{rh}}$~\cite{Bernal:2021qrl},  the red region is from the kinematical threshold $m_\phi>2m_\chi$.  From Fig. 7 we can further get that if inflatons decay process constitutes all the DM, the parameter $y_\chi$ should satisfies $2.081\times10^{-27}<y_\chi<5.294\times10^{-6}$.

Secondly, dark matter can be produced by the 2-to-2 scatter of  the inflaton. In this case~\cite{2to21,2to22,2to23}
\begin{equation}
\gamma=\frac{\pi^3 g_{*}^2}{3686400} \frac{T^{16}}{M_p^4 T_{\mathrm{rh}}^8} \frac{m_\chi^2}{m_\phi^2}\left(1-\frac{m_\chi^2}{m_\phi^2}\right)^{3 / 2},
\end{equation}
and the the corresponding DM yield is
\begin{equation}
\begin{aligned}
Y_0 & \simeq \frac{g_*^2}{81920 g_{*, s}} \sqrt{\frac{10}{g_{*}}}\left(\frac{T_{\mathrm{rh}}}{M_p}\right)^3\left[\left(\frac{T_{\mathrm{max}}}{T_{\mathrm{rh}}}\right)^4-1\right] \frac{m_\chi^2}{m_\phi^2}\left(1-\frac{m_\chi^2}{m_\phi^2}\right)^{3 / 2} \\
& \simeq 1.8 \times 10^{-2} \frac{T_{\mathrm{rh}} m_\chi^2}{M_p^{5 / 2} m_\phi^{1 / 2}}\left(1-\frac{m_\chi^2}{m_\phi^2}\right)^{3 / 2} .
\end{aligned}
\end{equation}
Combine with the upper bound of $T_{rh}<4.43\times10^{12}$GeV in section 6, we get that for reasonable values of $m_\chi$, the DM yield $Y_0$ is less than $10^{-18}$, the contribution to DM abundance is negligible.

Thirdly, dark matter can also be produced via the 2-to-2 scattering of SM particles, mediated by gravitons or inflatons.
For gravitons act as mediators, one can get the decay rate density is~\cite{2to24,2to25,2to26}
\begin{equation}
\gamma(T)=\alpha  \frac{T^8}{M_p^4},
\end{equation}
with $\alpha\simeq1.1\times10^{-3}$. The corresponding DM Yield through this channel is
\begin{equation}
Y_{ 0}=\left\{\begin{array}{lc}
\frac{45 \alpha}{2 \pi^3 g_{*, s}} \sqrt{\frac{10}{g_{*}}}\left(\frac{T_{r h}}{M_p}\right)^3, & \text { for } m_\chi \ll T_{r h}, \\
\frac{45 \alpha}{2 \pi^3 g_{*, s}} \sqrt{\frac{10}{g_{*}}} \frac{T_{r h}^7}{M_p^3 m_\chi^4}, & \text { for }T_{r h}  \ll m_\chi \ll  T_{\max }.
\end{array}\right.
\end{equation}
Similarly, if the mediators are inflatons, we can get
\begin{equation}
\gamma(T)\simeq \frac{y_{\chi}^2\lambda_{12}^2}{2\pi^5}\frac{T^6}{m_{\phi}^4},
\end{equation}
and the dark matter Yield is
\begin{equation}
Y_{ 0} \simeq \frac{135 y_\chi^2 \lambda_{12}^2}{4 \pi^8 g_{*, s}} \sqrt{\frac{10}{g_{*}}} \frac{M_p T_{r h}}{m_{\phi}^4}, \quad \text { for } T_{r h} \ll m_{\phi}.
\end{equation}
Using the upper bound of $y_\chi$, $\lambda_{12}$ and $T_{rh}$ in section 6, we carry out numerical calculation on $Y_0$, and for the graviton mediation the maximum value is on the order of $Y_0\sim10^{-23}$ for $m_\chi \ll T_{r h}$, $Y_0\sim10^{-25}$ for $T_{r h}  \ll m_\chi \ll  T_{\max }$, and for the inflatons as the mediators, the maximum value is $Y_0\sim10^{-48}$, which are all very small compared to the present DM density, so that we can ignore both cases of the 2-to-2 scattering processes.

Moreover, when the scalar perturbations corresponding to the peak of the power spectrum renter the horizon, it will produce the primordial black hole(PBHs) through gravitational collapse,  which could also be a candidate of DM \cite{Yokoyama:1995ex,GarciaBellido:1996qt,Clesse:2015wea,Garcia-Bellido:2016dkw,Cheng:2016qzb,Fu:2019ttf}. Thus we also calculate the abundance of PBHs using the Press-Schechter approach of gravitational collapse, and found that the peak mass of PBHs is around $0.7M_{\odot}$ and the fraction in dark matter is about $10^{-33}$, which is very small and can be negligible.

\section{Summary \label{sec:infl}}

In this paper we discuss the explanation of NANOGrav data using inflationary potential with double-inflection-point, and such potential can be realized by the polynomial potential from effective field theory with a cut off scale. For some choices of parameter sets, we analyze the inflation dynamics and show that the inflection point at the high scale predicts a scale-invariant power spectrum which consistent with the observations of the CMB. On the other hand, the inflection point at the low scale can cause an ultra-slow-roll stage,  which will generate a peak in the scalar power spectrum, the height of which is about $10^7$ magnitude of the CMB scale power spectrum.  When the perturbations corresponding to the peak value re-enters the horizon, it will induce GWs that can be detected by experiments. We calculate the energy spectrum of GWs and shown that the peak is at frequencies around nanohertz, which is within the frequency range of SKA and EPTA, and lies in the $2 \sigma$ uncertainty of the NANOGrav constraints. In addition, around millihertz, the  curves lies above the expected sensitivity curves of  ASTROD-GW, so it can be detected in near future.

After inflation ends, we assume that the inflaton is coupled with SM Higgs boson and singlet fermionic dark matter field. We analyze the reheating temperature, calculate the effect of one-loop CW corrections of the coupling terms, combined with the bounds of BBN, Lyman-$\alpha$, etc, we constraints the coupling parameters as $y_\chi<1.00687\times10^{-4}$ and $2.7451\times10^{-23}<\lambda_{12}/M_p<3.03996\times10^{-8}$. We also discuss the dark matter production  of inflaton decay, inflaton scattering and SM scattering, and find that the main way to produce dark matter is the direct decay of inflaton. If we assume that the inflaton decay process produce the whole DM abundance, the parameter $y_\chi$ should satisfies $2.081\times10^{-27}<y_\chi<5.294\times10^{-6}$.

\begin{acknowledgments}
This work was supported by ``the National Natural Science Foundation of China'' (NNSFC) with Grant No. 11705133,
and by ``the Natural Science Basis Research Plan in Shaanxi Province of China'' No. 2023-JC-YB-072.
\end{acknowledgments}

\end{document}